\documentstyle[preprint,aps,epsfig,amstex]{revtex} 
 
\tighten       

\begin{document} 
 
\draft

\title{Inelastic quantum transport in superlattices: success and 
failure of the Boltzmann equation} 
\author{Andreas Wacker} 
\address{Institut f{\"u}r Theoretische Physik,  
Technische Universit{\"a}t Berlin,  
Hardenbergstr. 36, 10623 Berlin, Germany} 
\author{Antti-Pekka Jauho} 
\address{Mikroelektronik Centret, Bldg 345 east, 
Danmarks Tekniske Universitet, 2800 Lyngby, Denmark} 
\author{Stephan Rott, Alexander Markus, Peter Binder, 
and Gottfried\ H.\ D{\"o}hler} 
\address{Institut f{\"u}r Technische Physik, Universit{\"a}t Erlangen, 
Erwin-Rommel-Str. 1, 
91058 Erlangen, Germany} 

\date{Physical Review Letters {\bf 83} (1999), 
sceduled tentatively for July, 26} 
\maketitle 
 
\begin{abstract} 
Electrical transport in semiconductor superlattices is studied within a
fully self-consistent quantum transport model
based on nonequilibrium Green functions, including phonon and impurity
scattering. We compute both the drift velocity-field relation and
the momentum  distribution function covering the whole field range
from linear response to negative differential conductivity. 
The quantum results are compared with the respective results 
obtained from a Monte Carlo solution of the Boltzmann equation.
Our analysis thus sets the limits of validity for the semiclassical 
theory in a nonlinear transport situation in the presence of inelastic 
scattering.
\end{abstract}
 
\narrowtext 
 
Quantum mechanical description of  electrical transport in strong electric  
fields is a notoriously difficult subject. 
As the distribution function of the electrons deviates strongly 
from equilibrium, the standard approaches of linear response 
do not apply. In some circumstances  
heated distribution functions may be useful,  
but in principle such an assumption should be 
justified by an underlying theory. 
Usually this problem of hot electrons is treated within the 
semiclassical Boltzmann transport equation (BTE) which can be solved 
to a desired degree of numerical accuracy by Monte-Carlo  
simulations (MC) \cite{JAC83}. This way both the distribution 
function and the current density can be obtained for arbitrary field 
strengths. 
Nevertheless one has to be aware of the severe assumptions  
implied by the use of the BTE. The electrons are viewed as 
classical particles with a dispersion relation given by the band structure 
and scattering is treated using Fermi's golden rule 
where energy conservation is strict. To recover the 
BTE from a quantum transport theory, such as  
nonequilibrium Green functions  (NGF)\cite{ZitNGF},  
several assumptions are 
required (see e.g. \cite{MAH90,HAU96}): (i) spectral functions with finite 
width are replaced by $\delta$-functions for free 
particles, (ii)  
the nonequilibrium Green function is assumed to be expressible 
in terms of the momentum distribution function, 
(iii) retardation effects 
are neglected to reproduce the Markovian  
collision integral \cite{BTEderivation}. Attempts 
to relax some of these assumptions have been made in earlier 
studies \cite{ZitREG,ZitANS}, and  
quantum corrections to distribution functions  
have been reported. Nevertheless we are not aware of any direct comparison 
of the results from BTE with a full quantum transport theory 
far from equilibrium, which is the task of this paper. 
 
Semiconductor superlattices (SL)\cite{GRA95d} provide a unique opportunity 
to study effects related to quantum transport because  
the width of the miniband can be tailored by the choice of 
the barrier and well widths as well as the material composition. 
For sufficiently large electric fields  negative differential 
conductivity (NDC) appears  which can be understood 
within the semiclassical theory by electrons traversing 
the whole miniband\cite{ESA70} thus performing  
Bloch-oscillations both in momentum and real space. 
This gives a distribution function which 
is far from any kind of thermal equilibrium. Many papers 
have analyzed the solution of the BTE in this miniband 
transport regime\cite{ZitMBT}. 
Alternatively, in the NDC region 
the electrical transport can be  
formulated in terms of Wannier-Stark hopping\cite{ZitWSH}. 
Finally, for small miniband width $\Delta$ and strong scattering,  
the transport can be viewed as a sequential tunneling 
process between adjacent wells\cite{ZitSQT}. 
 
While these simplified approaches have proven to be useful to analyze 
different experimental situations, it is clear that a complete 
description of transport in SLs requires a more sophisticated 
quantum mechanical treatment such as  density matrix theory  
\cite{SUR84,BRY97} or NGF \cite{LAI93}. 
Recently it was shown that a calculation based on NGF 
reproduces the results of the simplified approaches 
and determines their  
respective ranges of validity\cite{WAC98a}. 
All these approaches  employed a specific 
(heated) thermal distribution 
to model the in-scattering processes. 
While this can be a reasonable  approximation for the evaluation of averaged 
quantities such as the current density, such an  assumption is certainly not  
justified if the nonequilibrium electron distribution function  
itself is of interest. 
 
In this work we calculate the drift velocity-field relation as 
well as the electron distribution function both in a quantum transport 
model based on NGF and within the  BTE under stationary conditions. 
Identical system parameters and scattering matrix elements for 
impurity and phonon scattering are used.  
Our NGF calculation  provides us with a full self-consistent  
solution of the transport problem within 
the self-consistent Born-approximation for the scattering. 
(Thus, effects of higher order scattering such as 
weak localization are neglected.) This approach is similar 
to recent calculations for the resonant tunneling diode \cite{ZitLAK}. 
The BTE is solved by a MC-simulation. Its results 
can be easily interpreted in terms 
of standard concepts such as 
electron heating and  Bloch-oscillating electrons. We find 
excellent agreement between the drift velocities  
calculated semiclassically and quantum mechanically 
in the range of validity for the miniband transport 
discussed in Refs.~\cite{WAC98a,ROT99}. 
On the other hand, the failure of the BTE becomes obvious if 
the potential drop per period $eFd$ or the scattering width  
$\hbar/\tau_{\rm scatt}$ become larger than the miniband width $\Delta$. 
 
We consider a semiconductor superlattice with period $d$ 
and restrict ourselves to nearest neighbor  
coupling. The band structure reads  
$E(q,{\bf k})=2T_1 \cos(qd)+E_k$ with $T_1=-\Delta/4$ and 
$E_k=\hbar^2k^2/2m$. Here $q$ denotes the Bloch vector in the growth 
direction, which is restricted to $-\pi/d<q \le \pi/d$, and ${\bf k}$ 
is the two-dimensional Bloch vector perpendicular to the 
growth direction. $m=0.067m_e$ is the effective mass of the conduction band 
of GaAs. 
We restrict ourselves to the lowest miniband neglecting 
all phenomena related to intersubband processes  
(e.g., the second peak in the drift velocity-field relation at higher fields 
\cite{GRA95d,ZitSQT}). 
 
Three types of scattering processes are included: 
{\it Impurity scattering} at $\delta$-potentials  
with density $N_d$ (per unit area) and constant matrix element  
$V_{\rm imp}$  leading to a scattering rate  
$1/\tau_{\rm imp}=N_d\pi V_{\rm imp}^2\rho_0/\hbar$, where 
$\rho_0$ is the two-dimensional density of states.
(The impact of further elastic scattering processes like 
interface roughness can be taken into account by applying a reduced 
value of $\tau_{\rm imp}$.)
{\it Optical phonons} of 
energy $\hbar \omega_{\rm o}=36$ meV 
with a constant 
matrix element $M_{\rm o}$. 
We choose $M_{\rm o}$ such that the rate for spontaneous emission 
of a phonon (if allowed) is given by  
$1/\tau_{\rm o}=\pi M_{\rm o}^2\rho_0/\hbar=8$ ps$^{-1}$. These values 
are realistic for GaAs. 
In order to achieve energy relaxation for  energies lower 
than $\hbar \omega_{\rm o}$  we mimic  
{\it acoustic phonons} 
by a second phonon with constant energy  
$\hbar \omega_{\rm a}$. The ratio $\omega_{\rm o}/\omega_{\rm a}$ 
should be irrational to avoid spurious resonances and we choose 
$\hbar \omega_{\rm a}=\hbar \omega_{\rm o}(\sqrt{5}-1)/10 
\approx 4.4498$ meV.  The constant matrix 
element $M_{\rm a}$ was chosen to yield a scattering rate  
$1/\tau_{\rm a}=200$ ns$^{-1}$ \cite{COMmat}\nocite{ROT98}.  
These  matrix elements are used in the scattering term 
of the BTE assuming a thermal occupation  
$N_{\rm o/a}=[\exp(\hbar\omega_{\rm o/a}/k_BT)-1]^{-1}$ for the 
phonon modes with lattice temperature $T$. 
We solve the BTE by a MC procedure\cite{JAC83} and obtain the 
semiclassical distribution 
function $f_{\rm SC}(q,{\bf k})$ as well as the 
average drift velocity $v_{\bf drift}$ 
as a function of the electric field $F$ applied to the SL. 
 
Following Ref.~\cite{WAC98a} we use a basis of Wannier functions  
$\Psi(z-nd)$ (localized in well $n$) for the NGF 
calculation. The task is to evaluate  the retarded and lesser 
Green function $G^{\rm ret}_{m,n}(t,t',{\bf k})=-i\Theta(t-t') 
\langle \{a_{m}(t,{\bf k}),a_{n}^{\dag}(t',{\bf k})\}\rangle$ and 
$G^<_{m,n}(t,t',{\bf k})= 
i\langle a_{n}^{\dag}(t',{\bf k})a_{m}(t,{\bf k})\rangle$, respectively. 
Here $a_n^{\dag}(t,{\bf k})$ and $a_n(t,{\bf k})$ are the creation 
and annihilation operators for the state 
$\Psi(z-nd)e^{i({\bf k}\cdot {\bf r})}/A$ and  
$\{A,B\}$ denotes the anticommutator. 
Using the Dyson and Keldysh equation [Eqs.~(13,15) of Ref.~\cite{WAC98a}] 
the Green functions can be calculated for given self-energies  
$\tilde{\Sigma}^{\rm ret}_n$ and $\tilde{\Sigma}^{<}_n$  
(which are diagonal in the well index as short range scattering 
potentials are assumed). In Ref.~\cite{WAC98a} only elastic 
scattering was considered
and  an equilibrium approximation for 
$\tilde{\Sigma}^{<}_n$ was 
made to ensure energy relaxation. 
Here, instead, 
we calculate both $\tilde{\Sigma}^{\rm ret}_n$ and  
$\tilde{\Sigma}^{<}_n$  
self-consistently. For impurity scattering we use 
\begin{eqnarray} 
\tilde{\Sigma}^{{\rm ret}/<}_{n,{\rm imp}}({\cal E})&=& 
\frac{N_d}{A}\sum_{{\bf k}'}V_{\rm imp}^2 
G_{n,n}^{\rm ret/<}({\cal E},{\bf k}')\label{EqSigimp} 
\end{eqnarray} 
and for phonon scattering (see, e.g., Ch. 4.3 of Ref.~\cite{HAU96}) 
\widetext 
\begin{eqnarray} 
\tilde{\Sigma}^{<}_{n,{\rm o}}({\cal E})&=& 
\frac{|M_{\rm o}|^2}{A}\sum_{{\bf k}'}  
\left\{N_{\rm o}G^{<}_{n,n}({\cal E}-\hbar\omega_{\rm o},{\bf k}') 
+(N_{\rm o}+1)G^{<}_{n,n}({\cal E}+\hbar\omega_{\rm o},{\bf k}')\right\} 
\label{EqSiglessopt}\\ 
\tilde{\Sigma}^{\rm ret}_{n,{\rm o}}({\cal E})&=& 
\frac{|M_{\rm o}|^2}{A}\sum_{{\bf k}'} \Big\{ 
(N_{\rm o}+1)G^{\rm ret}_{n,n}({\cal E}-\hbar\omega_{\rm o},{\bf k}') 
+N_{\rm o}G^{\rm ret}_{n,n}({\cal E}+\hbar\omega_{\rm o},{\bf k}')\nonumber \\ 
&&\phantom{\frac{|M_{\rm o}|^2}{A}\sum_{{\bf k}'} \Big\{} 
+i\int\frac{d{\cal E}'}{2\pi}  
G^{<}_{n,n}({\cal E}-{\cal E}',{\bf k}') 
\left[\frac{1}{{\cal E}'-\hbar\omega_{\rm o}+i0^+}- 
\frac{1}{{\cal E}'+\hbar\omega_{\rm o}+i0^+}\right]\Big\}\, . 
\label{EqSigretopt} 
\end{eqnarray} 
\narrowtext 
In our numerical calculation we ignore the real part of the  
last  term 
containing the energy integral in $\tilde{\Sigma}^{\rm ret}_{n,{\rm o}}$,  
which  renormalizes the energy slightly.   
The contribution 
due to 'acoustic' phonons 
$\omega_{\rm a}$ is treated analogously. 
Note that the constant matrix elements  lead to 
{\bf k}-independent self-energies which implies a 
significant reduction in the computational effort. 
 
In our calculation we start with a guess for 
the self-energies, calculate the Green-functions and obtain a 
new set of self-energies from Eqs.~(\ref{EqSigimp}--\ref{EqSigretopt}). 
This procedure is continued iteratively until 
the new self-energies deviate by less than 0.5\% from the 
preceding ones. We checked the  convergence 
by comparing the currents starting from different initial guesses 
and found that even a stricter bound had to be used for 
low electric fields. Finally, we calculate the  
quantum momentum distribution function 
\begin{equation} 
f_{\rm QM}(q,{\bf k})=\frac{1}{2\pi i} 
\int d{\cal E}\sum_h e^{-ihqd}G^<_{h,0}({\cal E},{\bf k}) 
\end{equation} 
from the nondiagonal elements of $G^<$ and the current density 
\begin{equation} 
J(F)=\frac{2e}{(2\pi)^3}\int d^2k \int_{-\pi/d}^{\pi/d} dq 
 f_{\rm QM}(q,{\bf k}) 
\frac{1}{\hbar}\frac{\partial E(q,{\bf k})}{\partial q} 
\end{equation} 
which is equivalent to Eq.~(10) of Ref.~\cite{WAC98a}. 
 
\begin{figure} 
\noindent\epsfig{file=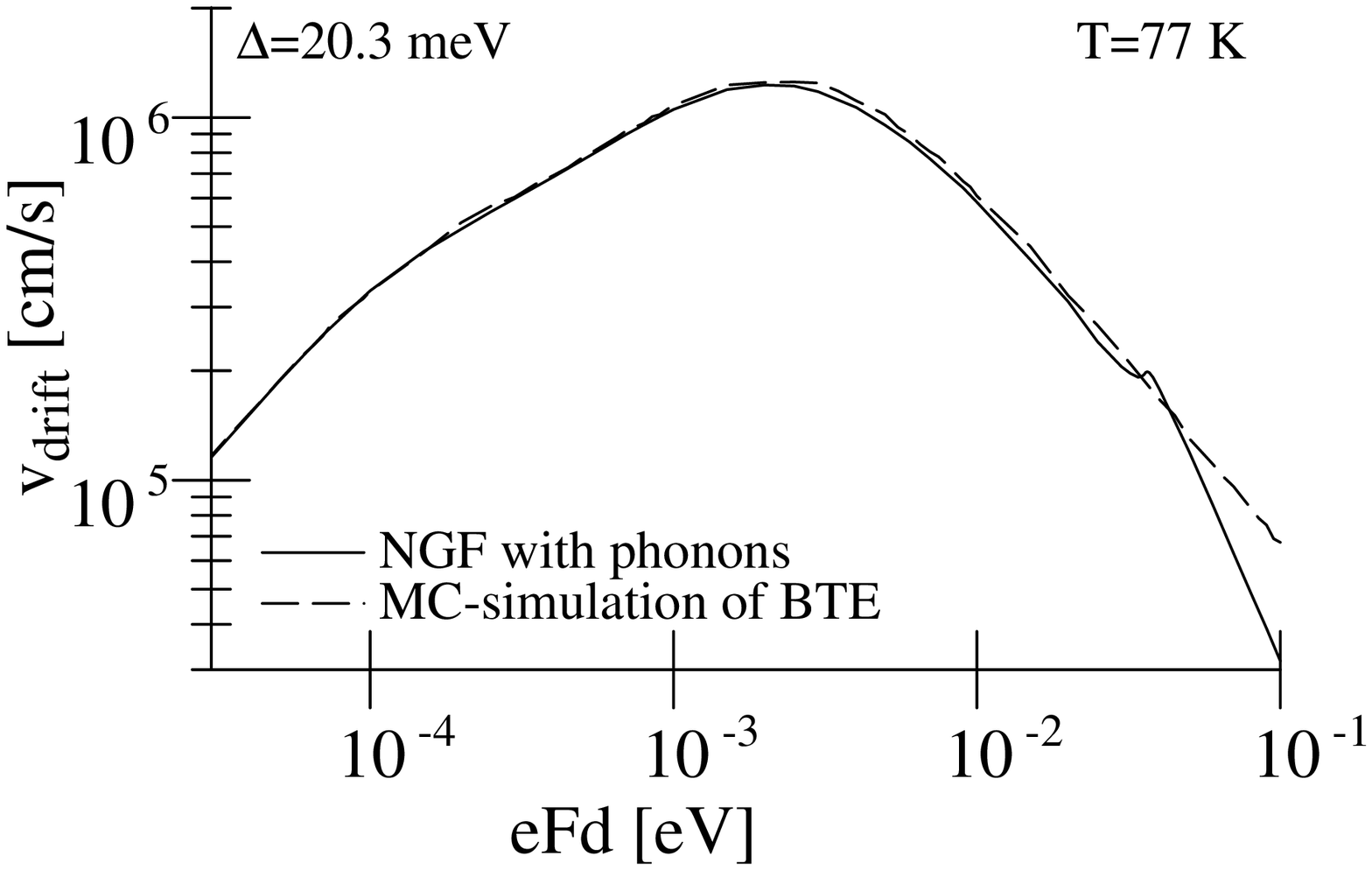,width=7cm}\\[0.2cm] 
\caption[a]{Drift-velocity versus field  
for a wide-band SL with 
$\Delta=20.3$ meV, $1/\tau_{\rm imp}=3$ ps$^{-1}$ for $T=77$ K.  
Full line: Calculation by nonequilibrium Green functions. Dashed line:  
MC-simulation of Boltzmann's transport equation. 
\label{Figkenn1}} 
\end{figure}

\begin{figure} 
\noindent\epsfig{file=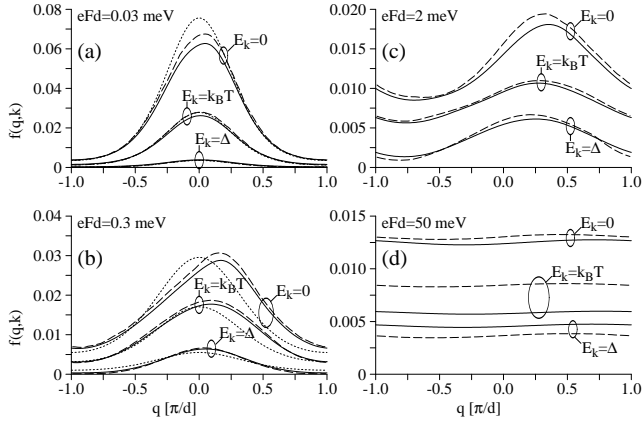,width=8.5cm}\\[0.2cm] 
\caption[a]{Electron distribution versus quasimomentum in the 
Brillouin zone of the miniband  for different values 
of $k$.  
(Parameters as Fig.~\ref{Figkenn1}) 
Full line: NGF calculation. Dashed line:  
MC-simulation of BTE. The dotted line shows the thermal 
distribution $\propto \exp[-E(q,{\bf k})/k_BT_e]$ with $T_e=T$ in (a)
and $T_e=140$ K in (b) for comparison.
\label{Figdistrq}} 
\end{figure}

\begin{figure} 
\noindent\epsfig{file=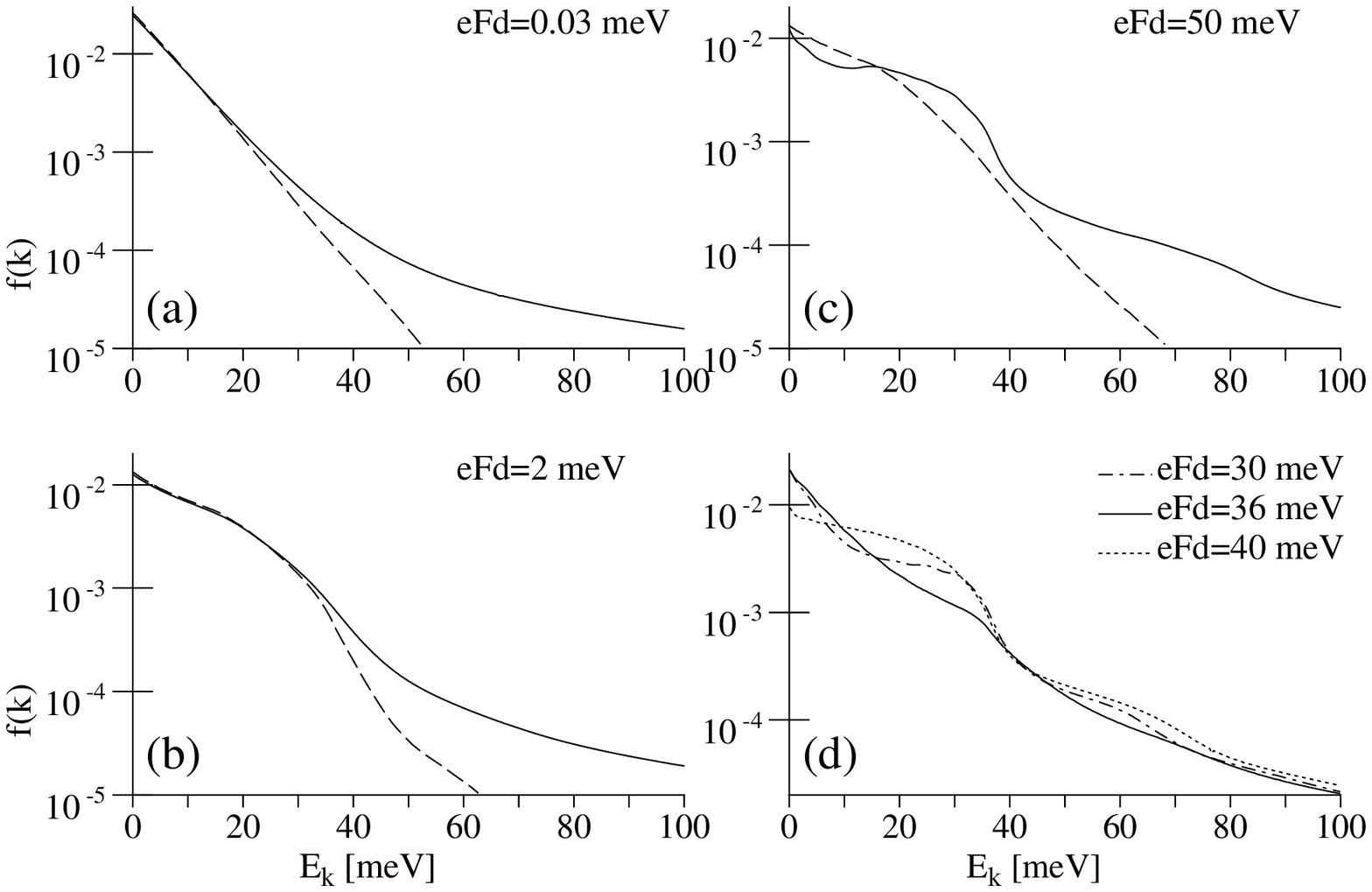,width=8.5cm}\\[0.2cm] 
\caption[a]{Electron distribution versus quasimomentum $k$ 
(Parameters as Fig.~\ref{Figkenn1}).  (a,b,c): Comparison  
between NGF calculation (full line) and BTE (dashed line); 
(d): Results from NGF for different fields. The BTE result (not shown) 
resembles the result from (c) for all three fields. 
\label{Figdistrk}} 
\end{figure}

{\em Results:} 
We first consider a wide-band superlattice 
with $\Delta=20.3$ meV and  period  $d=5.1$ nm. We 
assume an average carrier density of $10^{16}$ cm$^{-3}$ and $T=77$ K. The 
impurity scattering rate is taken to be $1/\tau_{\rm imp}=3$ps$^{-1}$. 
Fig.~\ref{Figkenn1} shows the calculated drift-velocity versus electric field. 
We find that the characteristics calculated from NGF and BTE 
are in excellent agreement for $eFd\lesssim\Delta/2$ as expected 
from the analysis of Ref.~\cite{WAC98a} because 
$\hbar/\tau\approx 2.3{\rm meV}\ll \Delta/2$. 
The shape of the $v_{\rm drift}(F)$ relation 
significantly deviates from the simple Esaki-Tsu shape 
$v_{\rm drift}\propto F/(F^2+F_c^2)$ with $F_c=\hbar/ed\tau$. A linear part is only observed for  
very low electric fields. Here the distribution function (see  
Fig.~\ref{Figdistrq}(a)) can be viewed as a distorted thermal 
equilibrium function. Thus the  
standard theory of linear response\cite{MAH90} makes sense 
yielding a linear part of the characteristics
(weak localization effects may affect the good agreement
between quantum transport and BTE at low temperatures).
In Fig.~\ref{Figdistrk}(a) we have shown the respective distribution 
versus $k=|{\bf k}|$, where  
$f(k)=d/(2\pi)\int_{-\pi/d}^{\pi/d} dq f(q,{\bf k})$. 
We find $f_{\rm SC}(k)\approx C\exp(-E_k/k_BT)$ from the       
BTE, where $C$ is a normalization constant.  
In the NGF calculation this behavior is only seen 
in the range $E_k\lesssim 20$ meV. For higher values of $E_k$ the quantum 
mechanical result is larger than the semiclassical one\cite{ZitREG} 
as energy broadening leads to the power-law 
$f_{\rm QM}(k)\sim C k_BT\Gamma/(2\pi E_k^2)$ in the momentum 
distribution function,  
where $\Gamma$ is the total scattering width. 
 
If the electric field increases, electron heating becomes important. 
For moderate fields the distribution function resembles a distorted  
equilibrium $f_{\rm eq}(q,{\bf k}) \propto \exp[-E(q,{\bf k})/k_BT_e]$  
with an increased electron temperature $T_e\approx 140$ K for 
$eFd=0.3$ meV,  see Fig.~\ref{Figdistrq}(b). This  
suppresses the mobility and causes  a   
sublinear increase of the current. Close to the maximum 
at $eFd=2$ meV the distribution function strongly deviates 
from any kind of equilibrium in $q$-space (see Fig.~\ref{Figdistrq}(c)), 
but the $k$-dependence can be still viewed as a heated  
distribution (see Fig.~\ref{Figdistrk}(b))  with $T_e\approx 190$ K  
for $E_k\lesssim \hbar \omega_{\rm o}$. 
The results look similar in the NDC region for $eFd=10$ meV 
(not shown here). In all cases (with $eFd\lesssim \Delta/2$) 
the distribution functions from BTE agree well with the  result 
from a full quantum mechanical description. Thus, the  
BTE can be viewed adequate in this parameter range. 
 
The situation chances dramatically for $eFd\gtrsim \Delta$, see 
Figs.~\ref{Figdistrq}(d),\ref{Figdistrk}(c). 
As the electrons can perform several 
Bloch-oscillations in the semiclassical picture, the distribution 
function is almost flat within the Brillouin zone of the miniband. 
The latter holds for the NGF result as well.  
However, the absolute 
values of the distribution functions differ significantly. 
The reason is the modification in scattering processes 
due to the presence of the electric field, leading to significant 
deviations in the distribution function. Therefore, also the drift velocities 
deviate significantly in this field region, see Fig.~\ref{Figkenn1}. 
Finally, note the phonon-resonance\cite{BRY97,ROT98}  
at $eFd=\hbar\omega_{\rm o}$ in the NGF result for the  
velocity-field relation. This feature cannot be recovered from the BTE where 
the field does not appear as an energy scale in the scattering term. 
The strong change of $f_{\rm QM}(k)$  close to the 
phonon resonance in the NGF calculation is shown in  
Fig.~\ref{Figdistrk}(d). In the high-field 
regions one typically encounters  patterns  
on the energy scales $\hbar \omega_{\rm o}$  and $eFd$ (due to the 
formation of the Wannier-Stark ladder) as well as differences and sums 
of both quantities. 
For $eFd=\hbar \omega_{\rm o}$ both energy scales coincide and
the distribution function does not show strong features seen at other 
high fields: this is due to the enhanced resonant 
tunneling from well to well which (i) leads to enhanced current (see Fig. 1) 
and (ii) prevents accumulation/depletion at certain energies, as seen at
non-resonant applied fields.
This shows that in the high-field region
the true distribution function cannot be approximated 
by a heated distribution, neither  
in $k$ nor in $q$-space. 
 
\begin{figure} 
\noindent\epsfig{file=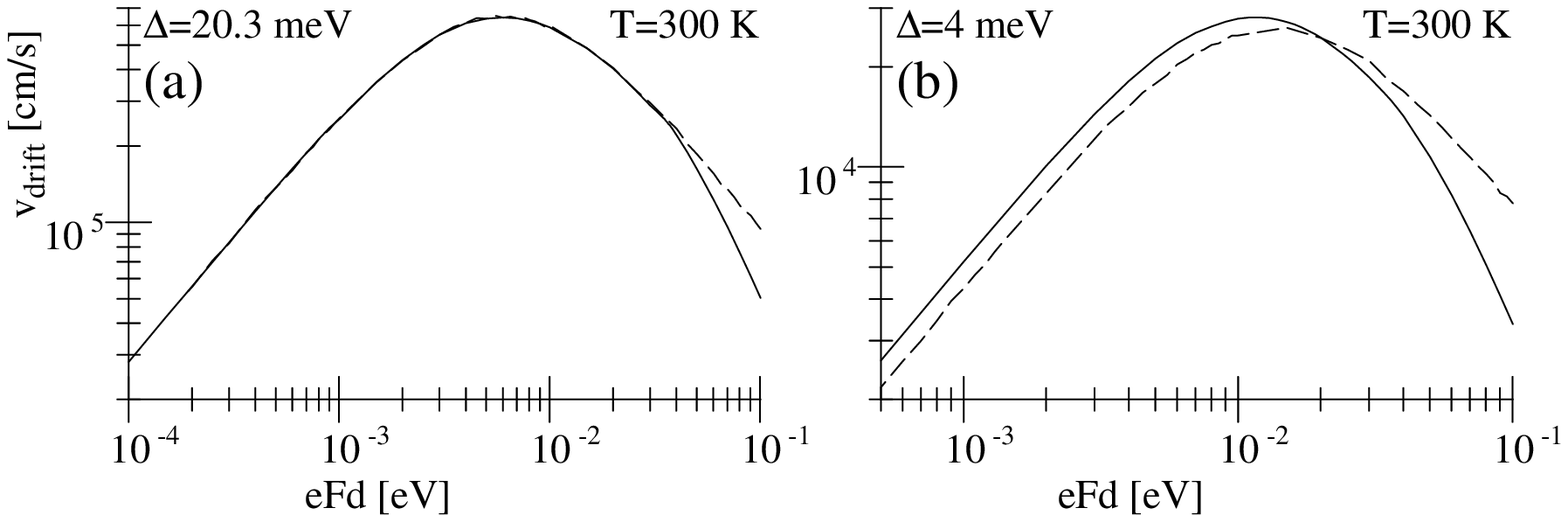,width=8.5cm}\\[0.2cm] 
\caption[a]{Drift-velocity versus field  for:  
(a) strongly coupled SL as in  Fig.~\ref{Figkenn1} but $T=300$K. 
(b)  weakly coupled SL with  
$\Delta=4$ meV, $1/\tau_{\rm imp}=15$ ps$^{-1}$, $T=300$K. 
Full line: NGF calculation. Dashed line:  
MC-simulation of BTE. 
\label{Figkenn2}} 
\end{figure}

For $T=300$ K electron heating is less important (Fig.~\ref{Figkenn2}a).  
Thus the linear regime extends to higher fields. Furthermore,  
the phonon-resonance is hardly visible.  Fig.~\ref{Figkenn2}(b) gives 
the result for a weakly coupled superlattice with $\Delta=4$ meV  
and an increased 
impurity scattering rate $1/\tau_{\rm imp}=15$ ps$^{-1}$. 
As $\hbar/\tau> k_BT,\Delta/2$  the calculated velocity-field relations 
deviate significantly both in the low-field mobility 
and the peak position. Again, the phonon-resonance is  
barely visible here, 
nor in the calculation for $T=77K$ where the deviation is even larger
(not shown here). 
  
{\em Conclusion:} We have performed a fully self-consistent 
quantum transport calculation for SLs which covers the whole range of
electric fields. A phonon resonance can be identified when
the potential drop per period equals the optical phonon energy.
The results for the drift-velocity and the 
quantum distribution function have been compared with data obtained 
from the semiclassical Boltzmann equation.
In wide-band SLs the Boltzmann equation gives reliable 
results concerning linear response at low fields, 
electron heating at moderate fields, and the onset of negative 
differential conductivity.  
In contrast, for high electric fields or weakly coupled 
SLs significant differences appear. 
In this case the quantum nature of transport is important and  
a semiclassical calculation may be seriously in error. 
We believe that an analysis of the kind presented above 
can be very useful in checking the quality of various approximation 
schemes.

\begin{thebibliography}{10}
\bibitem{COMmat} 
Note that a constant matrix element is equivalent to localized phonons, which 
  is far from realistic. Nevertheless our MC results with this approximation 
  resemble those with the correct matrix elements \cite{ROT98}. Thus our 
  approximations give at least qualitatively good results.
\end{thebibliography}
\end{document}